\documentclass{article}
\usepackage{spconf,amsmath,graphicx}
\usepackage[utf8]{inputenc} 
\usepackage[T1]{fontenc}    
\usepackage{hyperref}       
\usepackage{url}            
\usepackage{booktabs}       
\usepackage{amsfonts}       
\usepackage{nicefrac}       
\usepackage{microtype}      
\usepackage[dvipsnames]{xcolor}
\usepackage{placeins}
\usepackage{subcaption}
\usepackage{graphics}

\usepackage{enumitem}
\setlist{nosep, leftmargin=14pt}

\usepackage{mwe} 

\newcommand\figref{Figure~\ref}

\title{Learning Rotation Invariant Features for Cryogenic Electron Microscopy Image Reconstruction}
\name{Koby Bibas$^*$, Gili Weiss-Dicker$^*$, Dana Cohen, Noa Cahan, Hayit Greenspan \thanks{* Equal contribution}}
\address{Faculty of Engineering, Tel Aviv University, Israel}
\begin{document}
%
\maketitle
\begin{abstract}
Cryo-Electron Microscopy (Cryo-EM) is a Nobel prize-winning technology for determining the 3D structure of particles at near-atomic resolution. A fundamental step in the recovering of the 3D single-particle structure is to align its 2D projections; thus, the construction of a canonical representation with a fixed rotation angle is required. Most approaches use discrete clustering which fails to capture the continuous nature of image rotation, others suffer from low-quality image reconstruction. We propose a novel method that leverages the recent development in the generative adversarial networks. We introduce an encoder-decoder with a rotation angle classifier. In addition, we utilize a discriminator on the decoder output to minimize the reconstruction error. We demonstrate our approach with the Cryo-EM 5HDB and the rotated MNIST datasets showing substantial improvement over recent methods.
\end{abstract}

\section{Introduction}
Unsupervised feature learning algorithms have emerged as a promising tool for learning representations from data~\cite{bepler2019explicitly,caron2018deep,kaufman2019balancing}.
Learning invariant image representation enables machine learning algorithms to achieve good generalization performance while using a small number
of labeled training examples. 
An invariant representation is particularly valuable for the Cryo-EM, where the goal is to determine the 3D electron density of a particle from many noisy and randomly oriented 2D projections.
Having a model that aligns the 2D particle pose to a canonical predefined posture could 
significantly improved the 3D reconstruction of the particle~\cite{singer2020computational}.

Existing classic methods for the problem of determining the 3D structure of a particle use a Gaussian mixture model to group these 2D views. 
However, this assumes a discrete set of projections where it is known that particle conformations are continuous.
To face this issue,
more recent machine learning-based disentanglement approaches do not impose a specific structure on the learned latent representations~\cite{bepler2019explicitly}. These methods, however, use a variational autoencoder (VAE).
Using VAE induces blurriness to the reconstructed image which might eliminate key components in the particle structure.

In this work, we propose a different approach.
We use an encoder-decoder architecture and a discriminator. 
The discriminator penalizes the decoder generated images with rotated content.
In this way, we ensure that the generated content has a fixed orientation, which later can be utilized to reconstruct the 3D shape of the particle.
We demonstrate the effectiveness of our method on Cryo-EM and rotated MNIST datasets~\cite{bepler2019explicitly}. 
We improve the current leading approach mean squared error (MSE) by an order of magnitude on both the Cryo-EM 5HDB and the rotated MNIST datasets~\cite{bepler2019explicitly}.

\section{Related work}
In Cryo-EM, the main challenge is to determine the structure of a protein or a particle. 
In this section, we describe related work that tackles this problem.

Classic statistical methods assume that the many 2D projection observations of the particle arise from either a single structure or from a discrete mixture of structures. Assuming there are a finite number of possible projections, the particle views are grouped into a discrete number of clusters \cite{singer2020computational}. These conformations are confounded by orientation in the collected images. Thus, their goal is to cluster each projection image into one of the possible finite sets of projections.

Modern approaches tackle the problem of disentangling latent variables in an unconstrained setting~\cite{chen2016infogan}.
Others constrained the manifold of latent values to be homeomorphic to some known underlying data manifold to capture useful latent representations~\cite{falorsi2018explorations}.

The recently suggested \emph{spatial-VAE}~\cite{bepler2019explicitly} addresses this problem by formulating the generative model as a function of the spatial coordinates. This makes the reconstruction error differentiable with respect to latent rotation parameters which creates a representation that is independent of the content pose. 

A similar approach to ours was taken by the \emph{AttGAN}~\cite{he2019attgan}.
It uses an encoder-decoder architecture for facial attribute editing by conditioning the decoding of a given face latent representation on the desired attributes.
Notice that in our case the rotation angle is a continuous variable as opposed to face attributes and this imposes an additional challenge.

\begin{figure*}[tbh]
    \centering
     \includegraphics[width=0.95\textwidth]{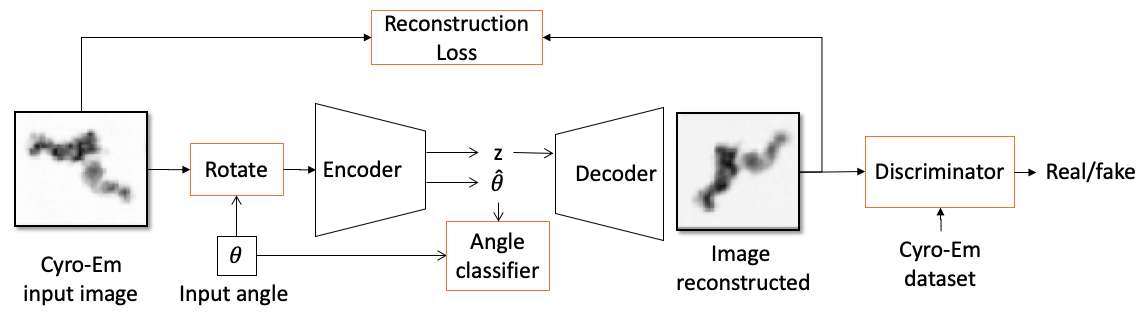}
    \caption{Our proposed scheme. We use encoder-decoder architecture with an angle classifier and a discriminator.}
    \label{fig:our_scheme}
\end{figure*}

\section{The proposed approach} \label{sec:method}
In this section, we describe our proposed method of disentangling the image content from the object pose in order to obtain a rotation-invariant image representation.

We employ an encoder-decoder architecture with a discriminator. 
A block diagram of the proposed system in shown in \figref{fig:our_scheme}. It contains the following stages: 
First, we apply a random rotation $\theta$ on the input image and propagate it through the encoder.
We utilize one scalar from the latent vector to represent the input image rotation.
Denote $\hat{\theta}$ as the predicted rotation (the latent variable value), we use the following loss for the angle classifier model
\begin{equation}
\mathcal{L}_{\textit{angle}}(\theta,\hat{\theta})= \exp\left\{|\theta - \hat{\theta}|\right\} - 1.
\end{equation}
When an accurate prediction is achieved, the $\exp$ value is 1 and the total loss is 0.
Next, we decode the latent vector with a decoder and compute the reconstruction loss with the original unrotated image.
Denote $x$ and $\hat{x}$ as the input image and the reconstructed image respectively, the reconstruction loss is
\begin{equation}
\mathcal{L}_{\textit{rec}}(x,\hat{x})= ||x - \hat{x}||_2 +  ||x - \hat{x}||_1.
\end{equation}

Finally, we use a discriminator that gets as input the reconstructed images and the given dataset.
We use the Wasserstein loss~\cite{arjovsky2017wasserstein}.
Denote $G$ as the generator (the decoder in our case), $D$ as the discriminator and $z$ as the latent vector without the rotation variable,
the loss functions of the discriminator and of the decoder are
\begin{equation}
\begin{split}
& \mathcal{L}_\textit{adv-disc}(x,D,G) = D(x) - D(G(z)), \\ 
& \mathcal{L}_\textit{adv-decoder}(x,D,G) = D(G(z)).    
\end{split}
\end{equation}
The adversarial loss is
\begin{equation}
\mathcal{L}_{\textit{adv}}(D,G) = \mathcal{L}_\textit{adv-disc}(D,G) + \mathcal{L}_\textit{adv-decoder}(D,G).
\end{equation}
The final training loss function is the combination of the above loss functions
\begin{equation}
\mathcal{L}(\theta,\hat{\theta},x,\hat{x}, D,G) = \mathcal{L}_{\textit{angle}}(\theta,\hat{\theta})
+ \mathcal{L}_{\textit{rec}}(x,\hat{x}) 
+ \mathcal{L}_{\textit{adv}}(D,G).
\end{equation}

In the following sections, we show that using our approach we manage to better reconstruct the unrotated images.  

\section{Datasets} \label{sec:datasets}
In order to evaluate performance, we conducted experiments using the following datasets.

\textbf{5HDB dataset~\cite{bepler2019explicitly,lin2016beta}.} A Cryo-EM dataset that contains simulated 2D projections with random rotations and additive random noise. 
The dataset includes 20K simulated projections of integrin $\alpha$-IIb with integrin $\beta$-3. The image size is 40x40. We used 16K and 4K images for training and testing respectively.

\textbf{Rotated MNIST~\cite{bepler2019explicitly}.} Each image from the MNIST dataset is rotated by a random angle sampled from $\mathcal{N}(0,\frac{\pi^2}{16})$.
Training and testing sets consist of 60K and 10K images respectively.

For both datasets, we normalized the pixel values such that their values is between 0 and 1.
We did not pre-process these datasets, besides the random rotation and normalization that were mentioned.

\section{Experiments} \label{sec:experimetns}
In this section, we present experiments\footnote{Code is available in \url{https://github.com/kobybibas/CryoEM_rotation_invariant}} that test our proposed encoder-decoder with a discriminator scheme as a method to disentangle the image content from the pose.
We compare our approach and the spatial-VAE method, which is considered a leading method in learning rotation-invariant features for Cryo-EM datasets.

For both 5HDB and rotated MNIST datasets, we trained our suggested model for 300 epochs with a learning rate value of $10^{-4}$ with a decrease by 0.1 after 200 epochs. For every 4 steps of the decoder, the discriminator was updated once. We used also a weight decay value that equals $10^{-5}$.

\begin{table}[tb]
    \centering
    \caption{Performance of our rotation invariant auto-encoder (AE) and the spatial-VAE model on the test set}
\resizebox{\columnwidth}{!}{%
    \begin{tabular}{cccc}
    \toprule
     Dataset & Method         & Average MSE &  Worst MSE \\
    \midrule
    5HDB    & Spatial-VAE    & 2.1    &  3.29  \\
            &  Rotation invariant AE     & 0.3    & 0.81  \\
    \midrule
    MNIST   & Spatial-VAE   & 66.07     &  121.82  \\
            &  Rotation invariant AE         & 0.02	    &  0.18  \\
    \bottomrule
    \end{tabular}
    \label{tab:5hdb_mse} \label{tab:mnist_mse}
}
\end{table} 

\begin{figure*}[tb]
    \centering
    \includegraphics[width=0.9\linewidth]{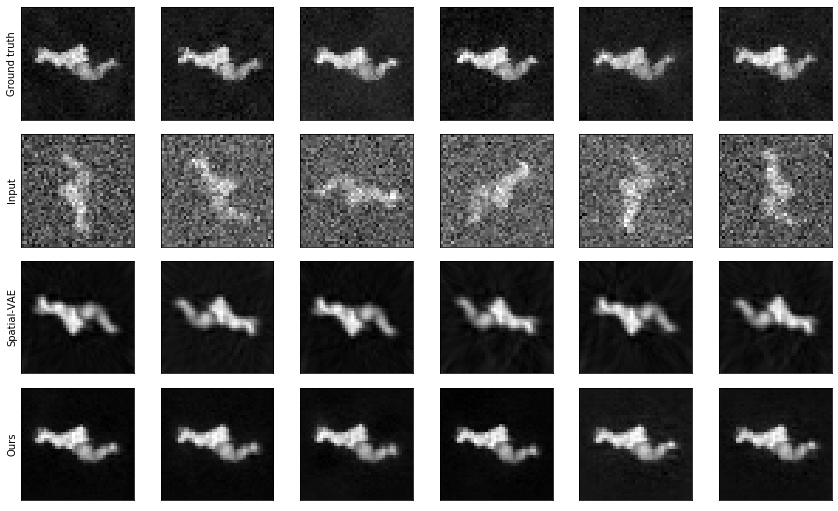}
        \caption{Proposed method results compared to baseline spatial-VAE for different rotation in the input. The first row is the ground true. The second row presents the model inputs. The 3rd and 4th rows show the spatial-VAE and our method results respectively.}
    \label{fig:5hdb_results}
\end{figure*}

\subsection{5HDB dataset}
In order to evaluate the performance of our method, we measured the MSE between the ground truth image and the decoder output.
The average MSE of the 5HDB test set is described in Table~\ref{tab:5hdb_mse}.
Our model outperforms the baseline in terms of average MSE compared to the original unrotated image by an order of magnitude. 

In \figref{fig:5hdb_results} we present the outputs of the compared models.
One can see that the object rotation of our model outputs is similar to the ground truth, where the rotation of the spatial-VAE model is different and is also changed based on the protein rotation in the input images: 
In the first, third, and fifth columns the orientation of the protein using the spatial-VAE model is flipped with respect to the ground truth.

We also evaluate the average MSE of the predicted angle by our method $|\theta-\hat{\theta}|^2$ on the 5HDB dataset. The result is an average MSE of 0.17 radians. 

\begin{figure}[tbh]
\centering
\centering
\includegraphics[width=1.0\linewidth]{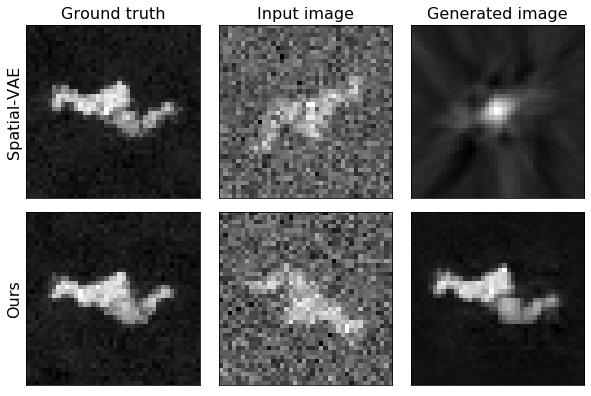}
\caption{5HDB dataset worst case MSE image. \label{fig:worstcase_5hdb}}
\end{figure}

The worst-case image is the image with the highest MSE between the ground truth image (the unrotated image with no noise) and the output of the model. 
A visualization of the worst-case 5HDB images of our model and the spatial-VAE model is presented in \figref{fig:worstcase_5hdb}. 
The spatial-VAE suffers from blurriness and did not reconstruct the image correctly. 
On the contrary, our model worst-case image can be considered as a successful reconstruction.

\begin{figure}[tbh]
    \centering
    \includegraphics[width=1.0\columnwidth]{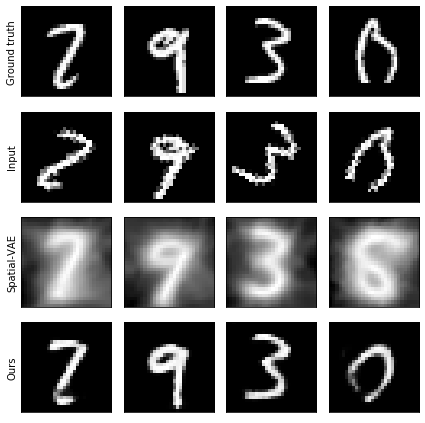}
        \caption{Rotated MNIST dataset. The first and  second rows are the ground truth and the input images respectively. The 3rd and 4th rows show the spatial-VAE and our method results.}
    \label{fig:mnist_results}
\end{figure}

\subsection{Rotated MNIST dataset}
We use the same metrics in the evaluation of the rotated MNIST dataset as in the 5HDB dataset.

The average MSE of the rotated MNIST test set is shown in Table~\ref{tab:mnist_mse}.
Our method attains an average MSE of 0.02 which is two orders of magnitude better than the spatial-VAE which has an average MSE of 66.07.

We show in \figref{fig:mnist_results} qualitative results of the rotated MNIST dataset.
As shown in the second the fourth columns, our method reconstructs the unrotated image with greater accuracy than the spatial-VAE method. 
In addition, the spatial-VAE outputs are blurred which explains the high MSE values of this method.

\section{Conclusion}
In this work, we suggested a novel encoder-decoder architecture with a discriminator to produce a canonical representation of cryogenic electron microscopy images.
Our suggested method offers an improvement on the 5HDB single-particle electron microscopy and rotated MNIST datasets. This is evident in both the quantitative and qualitative results. 

We are currently exploring additional variations to the proposed architecture, and its generalization to additional attributes. In the future, our method can be extended to additional modalities, such as CT and MRI imaging, and can help generate canonical representations and invariant reconstruction in various tasks.


\section{Compliance with Ethical Standards}
\label{sec:ethics}
This is a numerical simulation study for which no ethical approval was required.

\section{Acknowledgments}
\label{sec:acknowledgments}
No funding was received for conducting this study. The authors have no relevant financial or non-financial interests.

\bibliographystyle{IEEEbib}
\bibliography{references}

\end{document}